\documentclass[%
 showpacs,
 showkeys,
 amsmath,amssymb,
 aps,
 url,
 nofootinbib
]{revtex4-1}

\usepackage{amssymb}
\usepackage{amsmath}
\usepackage{graphicx}
\usepackage{geometry}

\begin{document}

\title{
Estimate of the Abelian $Z'$ decay width
}

\author{A.~Pevzner}
 \email{apevzner@omp.dp.ua}
\affiliation{Oles Honchar Dnipro National University, Theoretical Physics Department}

\begin{abstract}
The Abelian $Z'$ boson decay width is calculated in a model-independent approach. The analysis takes into consideration the special relations between the $Z'$ couplings to the Standard model fields proper to the renormalizable theories. The constraints on the $Z'$ couplings to the Standard model fermions corresponding to the narrow width approximation are derived. The results are compared with the benchmark $Z'$ models, such as $Z'_{\chi }$ and LR. It is shown that the current experimental limits on the Standard model Z width impose no limits on the $Z'$ width, so that the ratio $\Gamma _{Z'} / m_{Z'}$ is allowed to be up to 100\%. This fact is connected with the smallness of the $Z-Z'$ mixing. So, the finite-width $Z'$ is admitted and cannot be excluded from consideration. Also, the new constraints on the $Z-Z'$ mixing angle is derived which agrees with the corresponding earlier results of the LEP data analysis.
\end{abstract}

\keywords{Abelian $Z'$, new physics, decay width, LHC}

\maketitle

\section{Introduction}

The hypothetical heavy gauge $Z'$ boson is one of the most expected new particles in the Large hadron collider (LHC) experiments. Hundreds of possible scenarios of the Standard model (SM) extension are being discussed, and almost each of them contains a new neutral vector particle (\cite{Leike}, \cite{ZprEarlyLHC}). Thus the $Z'$ experimental discovery would open a great area for the new physics investigation.

The current experimental constraints on the $Z'$ mass are about $m_{Z'} >3.4$~TeV from the CMS and ATLAS data \cite{ReviewZPrimeSearches} with the 95\% confidence level (C.L.). Let us notice that, on the one hand, the model identification reach for the $Z'$ is about 5 TeV. This means that even if it is found above this threshold, it will be impossible to distinguish the specific basic model which it belongs to. On another hand, the LHC statistics for the events having invariant mass above 5 TeV is rather poor itself (for example, \cite{CMS_DirectSearch}, \cite{ATLAS_DirectSearch}). Such an effect origins from the behavior of the parton distribution functions (PDFs) that describe the internal proton structure. Hence we may roughly state that if the $Z'$ with the mass below 5 TeV is not discovered, it will not be discovered at the LHC at all. So, the current lower limit $m_{Z'} >3.4$~TeV is a good point to discuss the general possible further strategies of the $Z'$ searches.

The fact that no $Z'$ signal has been detected yet can be interpreted in two ways. The first and simple one is to suppose that the $Z'$ mass is really higher than 3.4 TeV and thus just has not been reached yet. Investigations of this case are not considered in the present paper. However, another possible explanation can be provided. It is usually assumed that the narrow width approximation (NWA) is applicable for the $Z'$ state, so that the peak of the $Z'$ production differential cross section lies within the experimental resolution of the LHC detectors. Usually, the NWA condition is understood as $\Gamma_{Z'}/m_{Z'}<$ several percent. This condition holds for the majority of the existing $Z'$ models, such as E6, LR \cite{Dittmar}, LH \cite{LH}, and others. Nevertheless, such an assumption is not obligatory. In the present paper, we show that in quite a general case, the finite-width $Z'$ state is not forbidden. This investigation is the main purpose of our work. In what follows, we will obtain the limits of applicability of the NWA and show in a model-independent approach that the ratio $\Gamma_{Z'}/m_{Z'}$ is allowed to be up to 100\%. 

\section{The model-independent approach}

We consider the Abelian $Z'$ boson, so that the SM electroweak gauge sector is extended with additional $U(1)$ group. Often, the $Z'$ searches are performed by using the specific models where different scenarios of the breaking Grand Unified Theory (GUT) symmetry group to the $U\left(1\right)$ are discussed. This approach leaves only the $Z'$ mass as a free parameter that must be fitted from the experiments. On the contrary, we use the model-independent description (\cite{Leike}, \cite{GulovSkalozub}) that allows to estimate not only the $Z'$ mass but also its couplings to the SM fields.

Provided that the $Z-Z'$ mixing takes place, we start from the effective Lagrangian
\begin{equation}
L_{Z\bar{f}f} =\bar{f}\gamma ^{\mu } \left[\left(v_{f}^{SM} -a_{f}^{SM} \gamma ^{5} \right)\cos \theta _{0} +\left(v_{f} -a_{f} \gamma ^{5} \right)\sin \theta _{0} \right]f\, Z_{\mu }, \label{eq:ZLagrangian}
\end{equation}
\begin{equation}
L_{Z'\bar{f}f} =\bar{f}\gamma ^{\mu } \left[\left(v_{f} -a_{f} \gamma ^{5} \right)\cos \theta _{0} -\left(v_{f}^{SM} -a_{f}^{SM} \gamma ^{5} \right)\sin \theta _{0} \right]f\, Z'_{\mu }, \label{eq:ZprLagrangian}
\end{equation}
where $v_{f} $ and $a_{f} $ are the vector and axial-vector couplings of the $Z$ and $Z'$ with the SM fermions, $\theta _{0} $ is the $Z-Z'$ mixing angle. So, the quantities $\left\{v_{f} ,a_{f} ,\theta _{0} ,m_{Z'} \right\}$ are unknown $Z'$ parameters that should be estimated.

However, without any additional conditions the number of the unknown parameters is too high, because we have vector and axial-vector couplings for each leptons and quarks generation. So, two assumptions are made here. The first one is that the theory is renormalizable. Also we suppose that the axial-vector constant $a_{f} $ is universal over all the fermions,
\begin{equation}
a\equiv a_{e} =-a_{\nu _{e} } =a_{d} =-a_{u} =... \label{eq:AUniversality}
\end{equation}
Under these assumptions, the following relations appear between the $Z'$ couplings\footnote{In the present paper, the coupling definitions are chosen accordingly to \cite{Leike}, not to \cite{GulovSkalozub}. They can be obtained from the definitions in \cite{GulovSkalozub} by replacements $a_f \to -2a_f$, $v_f \to 2v_f$.} \cite{GulovSkalozub},
\begin{equation}
v_{f} +a_{f} =v_{f_{*}} +a_{f_{*}} \quad \theta _{0} = 4a\frac{\sin \theta _{W} \cos \theta _{W} }{\sqrt{4\pi \alpha _{em} } } \left(\frac{m_{Z} }{m_{Z'} } \right)^{2} +O\left(\left(\frac{m_{Z} }{m_{Z'} } \right)^{4} \right), \label{eq:RGRelations}
\end{equation}
where $f$ and $f_{*}$ are the partners of the $SU\left(2\right)_{L} $ fermion doublet, $\alpha _{em} $ is the fine structure constant, $\theta _{W} $ is the Weinberg angle. Using of these relations reduces the number of the independent $Z'$ parameters drastically. Actually we have to estimate only $a$, $v_{e} $, and $v_{u} $ couplings at different values of $m_{Z'} $. In the following sections we will deal with these parameters only. Also we will use the ``normalized'' couplings
\begin{equation}
\bar{a}=\frac{1}{\sqrt{4\pi } } \frac{m_{Z} }{m_{Z'} } a, \quad \bar{v}_{e,u} =\frac{1}{\sqrt{4\pi } } \frac{m_{Z} }{m_{Z'} } v_{e,u} \label{eq:BarCouplings}
\end{equation}
The described model-independent approach covers many classes of the existing $Z'$ models. For example, in the popular $Z'_{^{\chi } } $and LR [6] models, the relations \eqref{eq:RGRelations} hold. In Tables I and II, the couplings of the $Z'$ belonging to these models are put.  All the values are normalized by the factor $e/\cos\theta_W$, so that the fermion current related to the $Z'$ is
\begin{equation}
j_{f}^{\mu } =\frac{e}{\cos \theta _{W} } \gamma ^{\mu } \left(\tilde{v}_{f} - \tilde{a}_{f} \gamma ^{5} \right). \label{eq:E6LRCouplings}
\end{equation}

\begin{table}[h!]
\label{tbl:ZChiCouplings}
\caption{The $Z'$ couplings in the $Z'_\chi$ model}
\centering
\begin{tabular}{|c|c|c|c|} \hline 
 & $e$ & $u$ & $d$ \\ \hline\hline
$\tilde{a}_{f} $ & $\frac{1}{2\sqrt{6} } $ & $-\frac{1}{2\sqrt{6} } $ & $\frac{1}{2\sqrt{6} } $ \\ \hline 
$\tilde{v}_{f} $ & $\frac{1}{\sqrt{6} } $ & 0 & $-\frac{1}{\sqrt{6} } $ \\ \hline 
\end{tabular}
\end{table}

\begin{table}[h!]
\label{tbl:LRCouplings}
\caption{The $Z'$ couplings in the LR model}
\begin{tabular}{|c|c|c|c|} \hline 
 & $e$ & $u$ & $d$ \\ \hline\hline
$\tilde{a}_{f} $ & $\frac{\alpha _{LR} }{4} $ & $-\frac{\alpha _{LR} }{4} $ & $\frac{\alpha _{LR} }{4} $ \\ \hline 
$\tilde{v}_{f} $ & $\frac{1}{2} \left(\frac{1}{\alpha _{LR} } -\frac{\alpha _{LR} }{2} \right)$ & $\frac{1}{2} \left(\frac{\alpha _{LR} }{2} -\frac{1}{3\alpha _{LR} } \right)$ & $-\frac{1}{2} \left(\frac{\alpha _{LR} }{2} +\frac{1}{3\alpha _{LR} } \right)$ \\ \hline 
\end{tabular}
\end{table}

The second, the third, and the fourth columns contain the $Z'$ couplings to the SM electron, up, and down quark respectively.

\section{The $Z'$ decay width: the NWA applicability}

\begin{figure}[h!]
\centering
\includegraphics{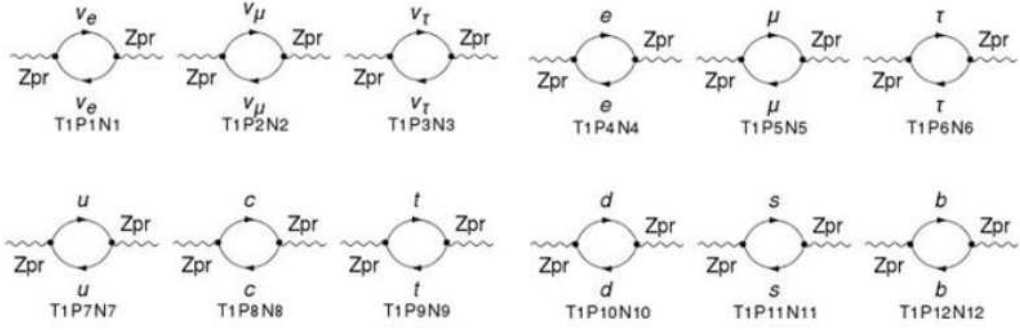}
\caption{The $Z'$ self-energy diagrams contributing to the $Z'$ decay width.}
\label{fig:ZPrWidthDiagrams}
\end{figure}

As discussed in the introduction, we are going to investigate the $Z'$ total decay width $\Gamma _{Z'} $. Using the Lagrangian \eqref{eq:ZprLagrangian} together with the relations \eqref{eq:RGRelations}, we calculate $\Gamma _{Z'} $ with the the optical theorem,
\begin{equation}
\Gamma _{Z'} =-\frac{\mathrm{Im}\left(\Pi _{Z'}^{^{\left(2\right)} } \left(m_{Z'}^{2} \right)\right)}{m_{Z'} } \label{eq:OpticalTheorem}
\end{equation}
where $\Pi _{Z'}^{^{\left(2\right)} } \left(k^{2} \right)$ is a $g_{\mu \nu } $-proportional part of the $Z'$ polarization operator in the second order by the electrical charge. It is assumed that the leptons and quarks of all the generations contribute to $\Pi _{Z'}^{^{\left(2\right)} } \left(k^{2} \right)$. In the Figure \ref{fig:ZPrWidthDiagrams}, all the diagrams corresponding to $\Pi _{Z'}^{^{\left(2\right)} } \left(k^{2} \right)$ are plotted using FeynArts (\cite{Mathematica}, \cite{FeynArts}).

Using \eqref{eq:ZprLagrangian} and \eqref{eq:OpticalTheorem}, we obtain the explicit expression for $\Gamma _{Z'} $. This allows us to establish the constraints on the $Z'$ couplings that correspond to the NWA. In the Figures~\ref{fig:ZPrWidthRegions1200GeV}~--~\ref{fig:ZPrWidthRegions5000GeV} below, we show some of the obtained coupling regions for $\Gamma_{Z'}/m_{Z'}<3\%$ and $\Gamma_{Z'}/m_{Z'}<10\%$ at different $Z'$ masses.

\begin{figure}[h!]
\centering
\includegraphics[scale=1.2]{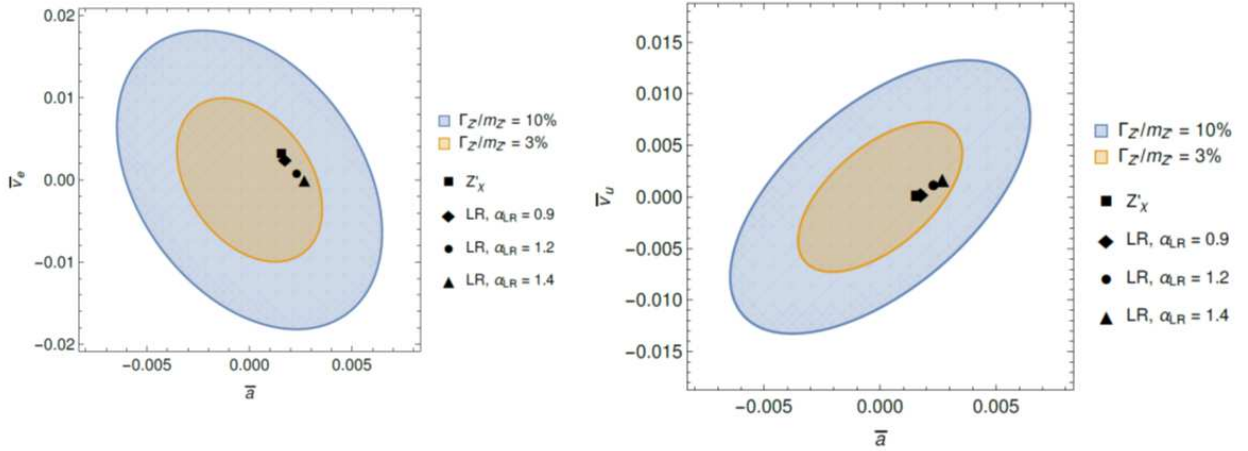}
\caption{The regions $\Gamma_{Z'}/m_{Z'}<3\%$ and $\Gamma_{Z'}/m_{Z'}<10\%$ at $m_{Z'} = 1.2$~TeV in $\left(\bar{a},\bar{v}_{e} \right)$ and $\left(\bar{a},\bar{v}_{u} \right)$ planes. The $Z'_{\chi } $ and LR models for different $\alpha _{LR} $ are marked.}
\label{fig:ZPrWidthRegions1200GeV}
\end{figure}

\begin{figure}[h!]
\centering
\includegraphics[scale=1.2]{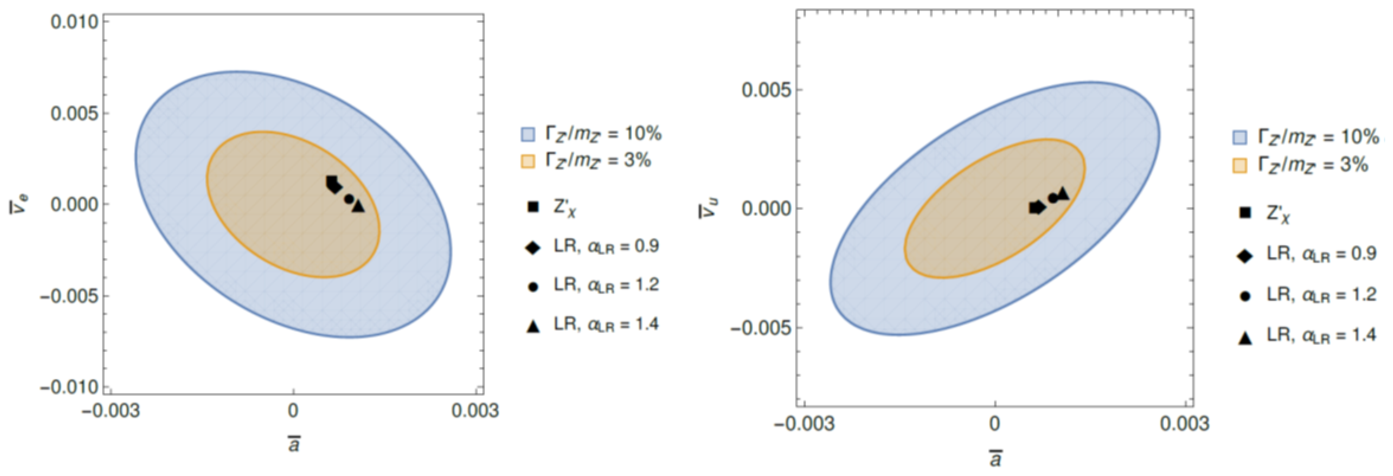}
\caption{The regions $\Gamma_{Z'}/m_{Z'}<3\%$ and $\Gamma_{Z'}/m_{Z'}<10\%$ at $m_{Z'} = 3$~TeV in $\left(\bar{a},\bar{v}_{e} \right)$ and $\left(\bar{a},\bar{v}_{u} \right)$ planes. The $Z'_{\chi } $ and LR models for different $\alpha _{LR} $ are marked.}
\label{fig:ZPrWidthRegions3000GeV}
\end{figure}

\begin{figure}[h!]
\centering
\includegraphics[scale=1.2]{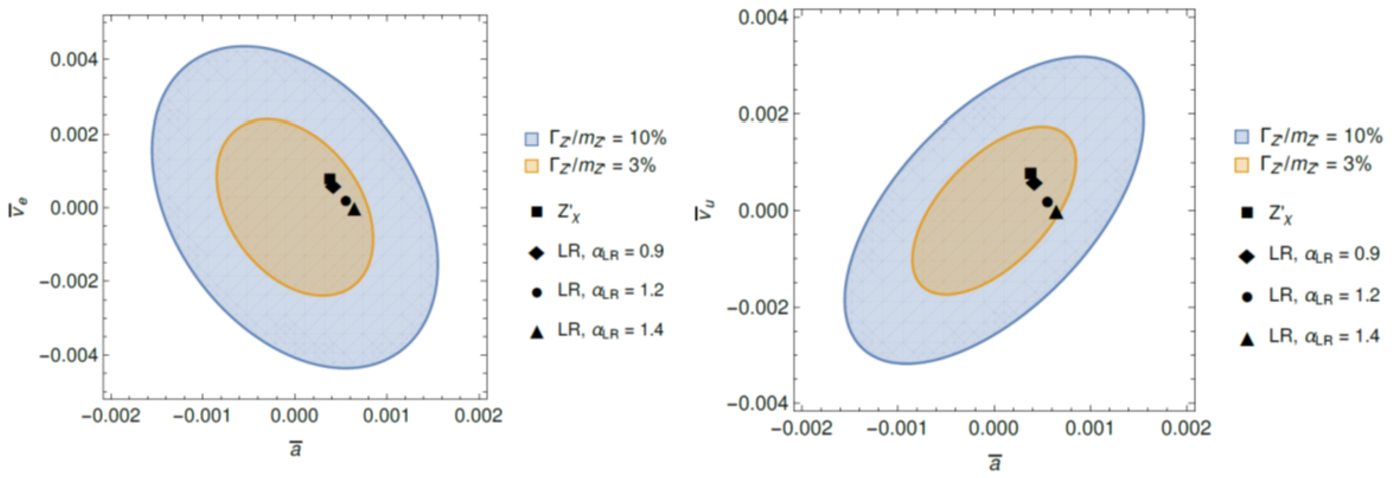}
\caption{The regions $\Gamma_{Z'}/m_{Z'}<3\%$ and $\Gamma_{Z'}/m_{Z'}<10\%$ at $m_{Z'} = 5$~TeV in $\left(\bar{a},\bar{v}_{e} \right)$ and $\left(\bar{a},\bar{v}_{u} \right)$ planes. The $Z'_{\chi } $ and LR models for different $\alpha _{LR} $ are marked.}
\label{fig:ZPrWidthRegions5000GeV}
\end{figure}

Thus we have established the regions of the $Z'$ couplings where the NWA is applicable.

\section{Analysis of the $Z'$ width from the Standard model $Z$ width constraints}

An important assumption being used for our model-independent approach is that the $Z'$ state is mixed with the $Z$ one. Hence the $Z'$ parameters can modify the Standard model $Z$ observables. At the same time, the latter were measured with the high precision at the LEP. So, when putting any constraints on the $Z'$ couplings, we have to check whether the $Z$ observables stay within their experimental uncertainties.

One of the well-measured $Z$ parameters is its decay width \cite{Z_PDG},
\begin{equation}
\Gamma _{Z} =\bar{\Gamma }_{Z} \pm \Delta \Gamma _{Z} =2.4952\pm 0.0023. \label{eq:ZWidth}
\end{equation}
In an extended theory,
\begin{equation}
\bar{\Gamma }_{Z} =\bar{\Gamma }_{Z} \left(a,v_{e} ,v_{u} ;m_{Z'} \right).
\end{equation}
To leave \eqref{eq:ZWidth} correct, we can vary the $Z'$ couplings in such the limits only that $\Gamma _{Z} $ is kept within the $\Delta \Gamma _{Z} $ interval. 

\begin{figure}[h!]
\centering
\includegraphics[scale=0.7]{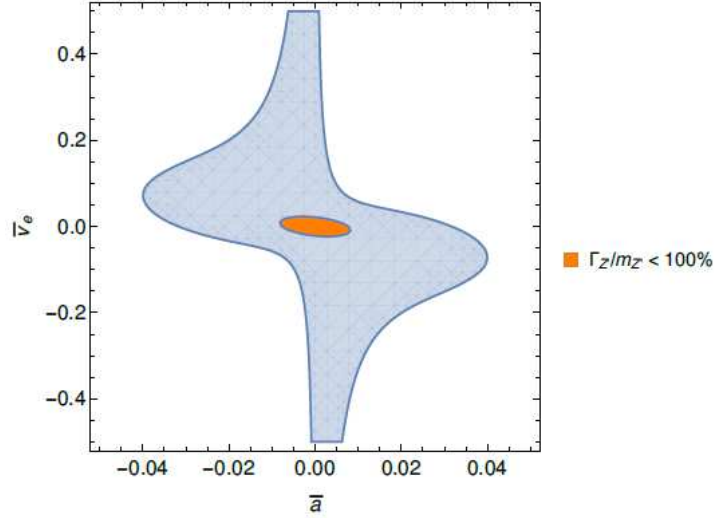}
\caption{The plot of the region \eqref{eq:ZprWidthEstimationRegion} in the $\left(\bar{a}, \bar{v}_e\right)$ plane.  The ellipse inside corresponds to the region $\Gamma_{Z'}/m_{Z'}<10\%$.}
\label{fig:ZprWidthEstimationRegion}
\end{figure}

Thus we establish the following estimation,
\begin{equation}
\left|\bar{\Gamma }_{Z} \left(a,v_{e} ,v_{u} ;m_{Z'} \right)-\bar{\Gamma }_{Z} \left(0,0,0;m_{Z'} \right)\right|\le \Delta \Gamma _{Z}, \label{eq:ZprWidthEstimationRegion}
\end{equation}
where $\bar{\Gamma }_{Z} \left(0,0,0;m_{Z'} \right)$ must obviously coincide with the SM value 2.4952~GeV. In the Figure \ref{fig:ZprWidthEstimationRegion}, the plot of the region \eqref{eq:ZprWidthEstimationRegion} is shown in the $\left(\bar{a},\bar{v}_{e} \right)$ plane.

The obtained results have several consequences. It can be seen from the figure that the axial-vector coupling $\bar{a}$ turns out to be constrained. So, we come to the new independent estimation on $\bar{a}$,
\begin{equation}
\left|\bar{a}\right|<0.04
\end{equation}
Using \eqref{eq:RGRelations}, we can also obtain the corresponding estimation on $\theta _{0} $,
\begin{equation}
\left|\theta _{0} \right|<10^{-3} -10^{-4},
\end{equation}
which agrees with the standard LEP constraints on the $Z-Z'$ mixing and some other known estimates \cite{AndreevPankov}.

An important conclusion is that the region \eqref{eq:ZprWidthEstimationRegion} admits any $Z'$ width,
\begin{equation}
\Gamma_{Z'}/m_{Z'}<100\%.
\end{equation}
This hereby means that the NWA is not a strictly necessary condition for the $Z'$ searches. This statement can be considered as a main result of our work.

\section{Summary and discussion}

The Abelian $Z'$ decay width has been analyzed in a model-independent approach which allows to estimate not only the $Z'$ mass but also the $Z'$ couplings to the SM fields. The investigation was performed by using the effective Lagrangian \eqref{eq:ZLagrangian}, \eqref{eq:ZprLagrangian}. The relations \eqref{eq:RGRelations} were used in order to decrease the number of the independent $Z'$ parameters. As a result, the $Z'$ coupling regions corresponding to the NWA have been derived. By investigating the $Z'$ influence on the $Z$ width it has been shown that the $Z'$ having any width up to $\Gamma_{Z'}/m_{Z'}<100\%$ is admitted. The fact that actually the $Z$ width measurements do not impose any additional restrictions on the $Z'$ width is nontrivial. It follows from our consideration that it is closely connected with the smallness of the $Z-Z'$ mixing angle. 

The obtained results mean that despite the fact that the NWA is a widely used and convenient assumption for the $Z'$ search, it should not hold obligatory. It follows from here that the standard direct search methods may occur irrelevant for the $Z'$.

For example, let us consider the $Z'$ with $m_{Z'} =4$~TeV and the couplings corresponding to the $Z'_{\chi } $ model. Let us believe it has a narrow peak $\Gamma_{Z'}/m_{Z'}=1\%$, so that $\Gamma _{Z'} =40$ GeV. (At this point we forget that the width is defined by the couplings just to illustrate inconsistency of the direct Z' searches for a finite-width Z' state.) Finally, imagine that we are searching for the $Z'$ in the Drell-Yan production at $\sqrt{s} =13$~TeV. Using the Lagrangian \eqref{eq:ZLagrangian}, \eqref{eq:ZprLagrangian}, we obtain that the $Z'$ production cross section integrated over the invariant mass bin $4000-40<M<4000+40$ GeV is about 0.05~fb. Taking the present CMS total integrated luminosity $L_{total} =\int L\,dt \approx 50\, fb^{-1} $, we conclude that it is possible to detect about $N_{Z'} =\sigma \, L_{total} \approx 2-3$ $Z'$ events (the acceptance effects are neglected).

Now let us take the finite-width case, for example, $\Gamma_{Z'}/m_{Z'}=30\%$. Then,$\Gamma _{Z'} =1200$~GeV. If we still believe that the $Z'$ has a narrow peak and search for it in the same invariant mass bin $4000-40<M<4000+40$~GeV, we obtain that the $Z'$ production cross section is about $10^{-4} $~fb. In this case $N_{Z'} \ll 1$, thus no $Z'$ signal will be detected at the LHC.

It follows from the above that the indirect off-peak approaches should be used. For example, such the methods have been used successfully in \cite{PevznerSkalozub} to estimate the $Z'$ couplings from the forward-backward asymmetry of the Drell-Yan process at low energies. The necessity of the finite-width $Z'$ investigations is also emphasized in some recent papers (for example, see \cite{AccomandoWidth}). Our future works will be devoted to the further development of the indirect $Z'$ detection.

\begin{acknowledgements}
The author is grateful to V.~Skalozub, A.~Pankov, and A.~Gulov for the fruitful discussions. The work is carried out within the international cooperation program between Ukraine and CERN. Also, the author acknowledges the receipt of the grant from the Abdus Salam International Centre for Theoretical Physics, Trieste, Italy.
\end{acknowledgements}

\bibliography{Pevzner_Zpr_Width_2017_for_TeX}

\end{document}